\documentclass[journal]{IEEEtran}

\normalsize

\ifCLASSINFOpdf
\else
\fi

\usepackage[font=footnotesize]{subfig}
\usepackage{array}
\usepackage{fixltx2e}
\usepackage[cmex10]{amsmath}
\usepackage[font=footnotesize, margin=0.0cm]{caption}
\usepackage{graphicx}
\usepackage{amsmath}
\usepackage{amsthm, amssymb}
\usepackage[overload]{empheq}
\usepackage{commath}
\usepackage{caption}
\usepackage{multirow}
\newcommand{\RN}[1]{%
  \textup{\uppercase\expandafter{\romannumeral#1}}%
  }
\usepackage{mathtools}

\DeclareMathAlphabet{\mathpzc}{OT1}{pzc}{m}{it}
\usepackage{cite}
\usepackage{color}
\usepackage{subfig}
\usepackage{url}

\graphicspath{{../fig/}}   

\begin{document}

\title{Reconfigurable Intelligent Surface Aided \\Mobile Edge Computing}

\author{Tong Bai,~\IEEEmembership{Member,~IEEE},
			Cunhua Pan,~\IEEEmembership{Member,~IEEE},\\
			Chao Han,~\IEEEmembership{Student Member,~IEEE},
			and Lajos Hanzo,~\IEEEmembership{Fellow,~IEEE}
\thanks{T. Bai is with the School of Cyber Science and Technology, Beihang University, Beijing 100191, China. (e-mail: t.bai@outlook.com).}
\thanks{C. Pan is with the School of Electronic Engineering and Computer Science, Queen Mary University of London, London, E1 4NS, U.K. (e-mail: c.pan@qmul.ac.uk)}
\thanks{C. Han is with the School of Electronic and Information Engineering, Beihang University, Beijing 100191, China (e-mail: chaohan@buaa.edu.cn).}
\thanks{L. Hanzo is with the School of Eletronics and Computer science, University of Southampton, SO17 1BJ, UK.  (e-mail: lh@soton.ecs.ac.uk)}
}

\markboth{IEEE Draft}%
{Submitted paper}

\maketitle

\begin{abstract}
Given the proliferation of wireless sensors and smart mobile devices, an explosive escalation of the volume of data is anticipated. However, restricted by their limited physical sizes and low manufacturing costs, these wireless devices tend to have limited computational capabilities and battery lives. To overcome this limitation, wireless devices may offload their computational tasks to the nearby computing nodes at the network edge in mobile edge computing (MEC). At the time of writing, the benefits of MEC systems have not been fully exploited, predominately because the computation offloading link is still far from perfect. In this article, we propose to enhance MEC systems by exploiting the emerging technique of reconfigurable intelligent surfaces (RIS), which are capable of `reconfiguring' the wireless propagation environments, hence enhancing the offloading links. The benefits of RISs can be maximized by jointly optimizing both the RISs as well as the communications and computing resource allocations of MEC systems. Unfortunately, this joint optimization imposes new research challenges on the system design. Against this background, this article provides an overview of RIS-assisted MEC systems and highlights their four use cases as well as their design challenges and solutions. Finally, their performance is characterized with the aid of a specific case study, followed by a range of future research ideas.
\end{abstract}

\IEEEpeerreviewmaketitle

\section{Introduction}

In the forthcoming Internet-of-Things (IoT) era, myriads of machines, sensors, and electronics gadget will be connected through the Internet \cite{hong2020space}. By analyzing the data collected, the IoT concept facilitates the monitoring and control of the physical world in diverse scenarios, including, automated home appliances, smart communities, autonomous driving, intelligent transportation, industrial automation, and emergency management. This innovative paradigm facilitates a novel interaction pattern among ``things" and humans, and will substantially reshape our daily lives.
The Softbank Group has predicted that a trillion devices will be connected to the Internet and will create $11$ trillion US dollar in value by $2025$. 

\begin{figure*}[t!]
\centering
\includegraphics[width=6.5in]{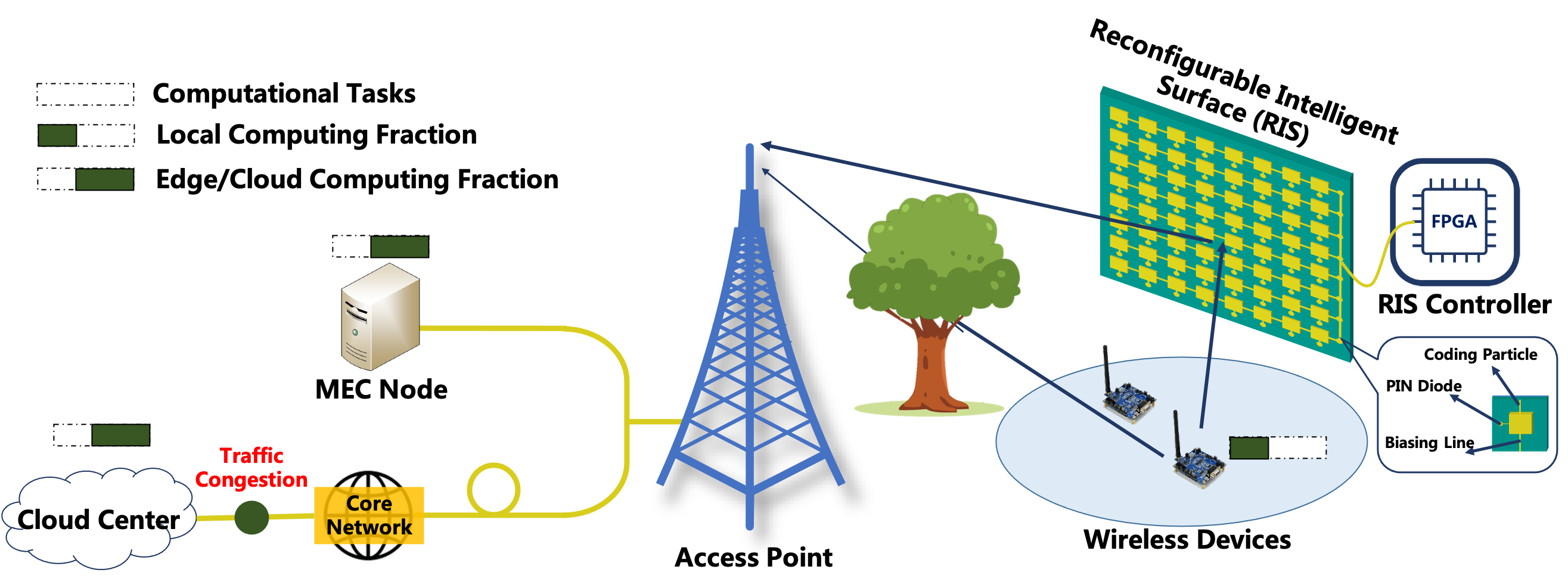}
\caption{Illustration of a reconfigurable intelligent surface (RIS)-assisted mobile edge/cloud computing (MEC) system, where a fraction of computational tasks are offloaded from wireless devices to the mobile edge computing node via an access point (AP) with the aid of a RIS, for enabling edge computing or both local computing and edge computing. RIS typically comprises a field-programmable gate array (FPGA) controller and a large number of reflection elements. Here PIN refers to positive-intrinsic-negative.}
\label{fig:RIS_MEC}
\end{figure*}

In contrast to the conventional cloud computing system where devices tend to act as data sinks, for example, downloading a Netflix video clip, the devices in the IoT era become heterogeneous and immense volumes of data are likely to be produced. For example, a raw data rate of $5.6~\rm{Gb/s}$ is required for a virtual reality handset having a $2160 \times 1200$ resolution \cite{han2019mobile}, while Instagram users post nearly $220,000$ new photos every single minute. As shown in Fig.~\ref{fig:RIS_MEC}, if all raw data were offloaded to the cloud, it would cause excessive downloading latency due to the long end-to-end delay and severe traffic congestion in the core network. For reducing this downloading delay, the mobile edge computing (MEC) paradigm may be harvested for extending the functionality of conventional cloud computing towards the edge of the network. Here ``edge" is defined as the computing, storage, and networking resources along the path between the data sources and cloud data centers. For example, the gateway in a smart home constitutes the edge between the home appliances and the cloud, while a `cloudlet' is the edge between mobile devices and the cloud. With the assistance of MEC nodes, the raw data can be offloaded to MEC nodes and then processed or stored locally for avoiding fetching them from the distant cloud via the core network, since the latter increases the delay. This regime tends to substantially relieve the traffic congestion of the core network and eliminate the end-to-end delay along the path between the edge and the cloud, hence it is especially beneficial for delay-sensitive applications. 

At the time of writing, however, the potential of MEC systems has not been fully exploited, predominantly because the computation offloading link is far from perfect. More explicitly, the devices at the cell edge or those behind line-of-sight (LoS) blockages usually suffer from low offloading rates, which increases both the latency and the energy consumption of computation offloading. The resultant reduced offloading probability also implies that a large fraction of the computing resources available at the MEC servers have to be idle due to the limited volume of tasks received, which limits the overall benefits of MEC systems. Furthermore, since the data offloaded by the devices usually contains private nature, its security should also be guaranteed. Therefore, it is imperative to improve the efficiency of MEC systems by enhancing their offloading links.

To this end, Reconfigurable Intelligent Surfaces (RIS) come to rescue for enhancing the offloading links \cite{huang}. Explicitly, RIS comprises of a large number of low-cost passive reflecting elements, each of which is capable of beneficially adjusting the phase and  possibly even the amplitude of the reflected signal, for improving the signal propagation environment. 
Specifically, if the direct LoS link between the wireless devices and MEC nodes is blocked by obstacles, the data can be offloaded via the RIS-aided reflected link. Explicitly, reflection-based beamforming can be realized by jointly optimizing the reflection coefficients of RISs, both for enhancing the offloading rate of the devices at the cell edge and for improving the physical-layer security (PLS). Thus, the overall performance of MEC can be substantially improved with the aid of RISs. Furthermore, in contrast to conventional transceivers, RISs require no high-power radio-frequency (RF) chain, only low-complexity control circuits. Hence, they can be densely deployed at a low cost and low energy consumption, for supporting ubiquitous MEC. Finally, embedding RISs into the existing MEC systems does not necessitate new protocols and/or MEC-hardware.

The rest of this paper is organized as follows. In Section~\RN{2}, we detail the RIS-assisted MEC system and discuss four use-cases as well as its research challenges. The beneficial role of RISs in MEC systems is presented with the aid of a case-study in Section~\RN{3}. In Section~\RN{4}, we highlight a number of future research opportunities and conclude in Section~\RN{5}.

\section{Reconfigurable Intelligent Surface Empowering Mobile Edge Computing}

\subsection{Fundamentals of RIS}

The RIS, as shown in Fig.~\ref{fig:RIS_MEC}, also known as an intelligent reflecting surface (IRS), is a software-controlled meta-material surface that has a programmable electromagnetic response \cite{cui2014coding}. The phase shift and possibly even the amplitude of the incident signals can be adjusted by each reflective element, by using positive-intrinsic-negative (PIN) diodes, or field-effect transistors (FETs) or alternatively micro-electromechanical systems (MEMS). Upon adopting a field-programmable gate array (FPGA) based micro-controller, the reflection elements' responses can be concurrently adjusted by controlling their switch state and direct current bias voltages. As such, we may establish a controllable communications link, where the wireless propagation environment can be adapted in a near real-time manner.

The advantages of RISs over other alternatives, namely full-duplex relays \cite{di2020reconfigurable} and active metasurfaces \cite{hu2018beyond}, are now detailed as follows. Full-duplex relays \cite{di2020reconfigurable} transmit and receive simultaneously. Hence they inevitably suffer from self-interference (SI). By contrast, the passive RISs are completely free from SI, hence eliminating both the SI cancellation hardware and the corresponding latency. More explicitly, the relays rely on both digital-to-analog converters (DACs) and analog-to-digital converters (ADCs), mixers, low-noise amplifiers, as well as other active electronic components, which inevitably impose much higher hardware complexity.
Active metasurfaces \cite{hu2018beyond} are capable of providing exceptional control of signals, however they have high complexity and excessive power consumption. In a nutshell, RISs have the advantage of high energy efficiency and low computational complexity. Hence they are eminently suitable for MEC systems.

\subsection{RIS-Assisted MEC Systems}

Fig.~\ref{fig:RIS_MEC} illustrates a RIS-assisted MEC system, which is comprised of an access point (AP) connected to the edge computing node using high-speed optical fiber, a RIS, and a number of wireless devices, each of which has specific computational tasks to be processed. If a wireless device is incapable of processing all of these tasks due to its limited power supply and/or restricted computational capability, some fraction of it or possibly all the tasks can be offloaded to the MEC node via the wireless AP and the concatenated fiber link of Fig.~\ref{fig:RIS_MEC}. However, the direct links between these devices and the AP may be blocked by obstacles. In this scenario, their computation offloading may mainly rely on the RIS-aided reflected link.
The MEC node is equipped with an edge manager, used for making scheduling decisions concerning the computational tasks to be offloaded from the wireless devices and for allocating both the computational resources at the MEC node and the communications resources for each wireless device, according to both the near-instantaneous channel and computational load information, both the wireless devices and of the MEC node, as well as the quality of experience (QoE) required by the specific applications. 

\begin{figure*}[t!]
\centering
\includegraphics[width=6.5in]{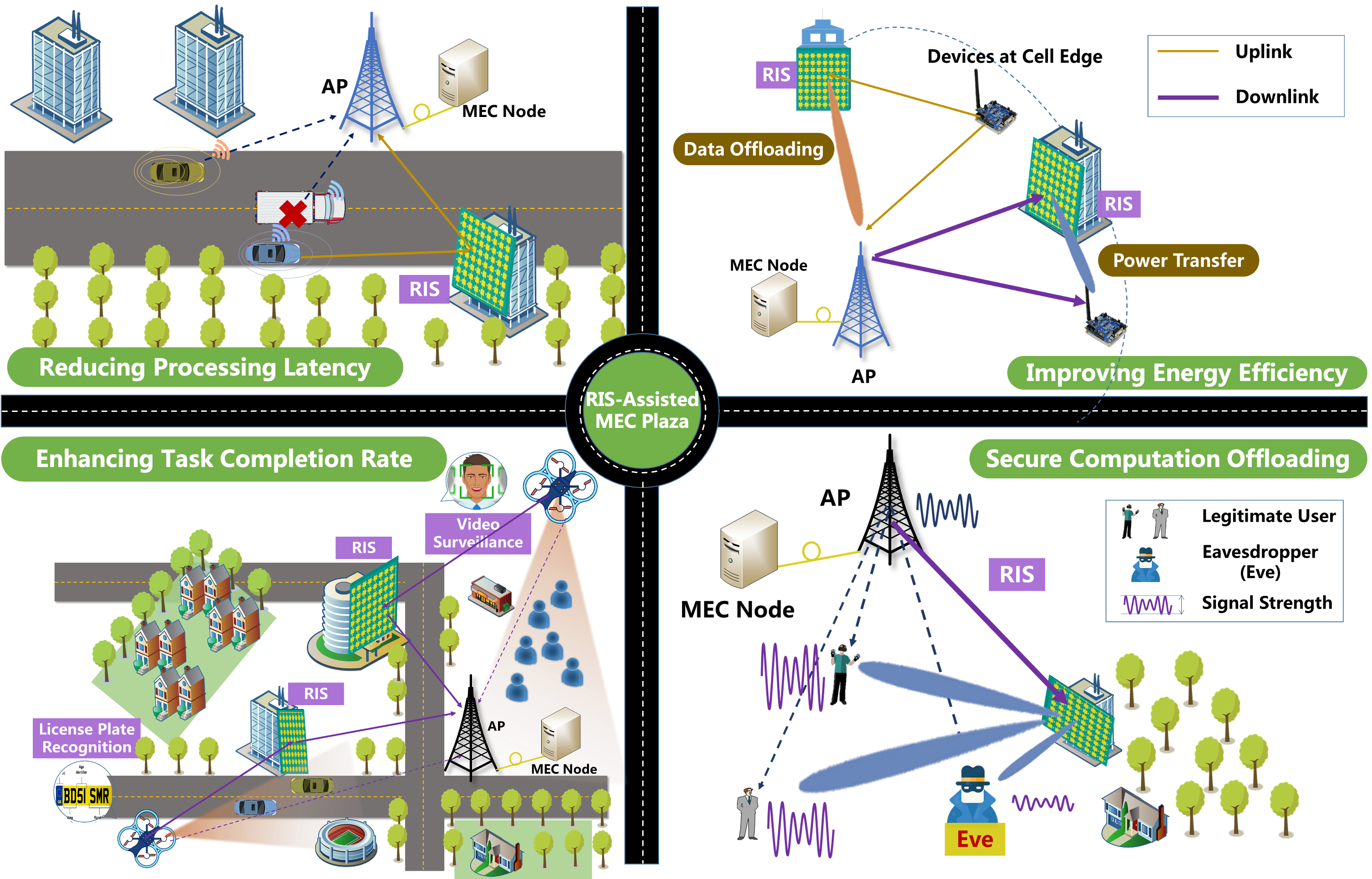}
\caption{Use Cases of reconfigurable intelligent surface (RIS)-assisted mobile edge computing (MEC) systems, including shortening processing latency, improving energy efficiency, enhancing the task-completion rate, and secure computation offloading.}
\label{fig:use_case}
\end{figure*}

In contrast to the conventional MEC system having no RISs, the employment of RISs provides a new degree-of-freedom. Specifically, the reflection coefficients of RISs can be jointly designed both with the communications and with computing resource allocation for meeting the stringent requirements of emerging applications.

\subsection{Use Cases}

Fig.~\ref{fig:use_case} highlights four potential use cases of RISs in MEC systems, which are detailed as follows.

\subsubsection{Reducing Processing Latency} 

Low latency is required by a variety of applications, such as augmented/virtual reality, online gaming, remote desktop, health care, and connected vehicles. The processing latency in MEC systems is the sum of the delay induced by computational offloading, by computing at the MEC nodes, and by feeding the computational results back. Given that the computational result is typically of a limited size, the feedback latency may be deemed negligible. Hence, the latency of MEC systems is predominantly owing to the first two components, which can be substantially reduced upon utilizing RISs \cite{bai2020latency}. To elaborate a little further using an example of collective environmental awareness in intelligent transportation systems, both the video and still images gleaned by vehicles have to be processed collaboratively by the MEC nodes in a near real-time manner for driver assistance and traffic management, because the vehicles usually do have sufficient computing power for carrying out their own tasks, but they are unable to have access to the scene observed by the other vehicles.
However, the vehicles' LoS computational offloading links are often blocked by trucks, buses, etc. As shown in Fig.~\ref{fig:use_case}, RISs can be harnessed for establishing a virtual link based on their reflection, which may substantially enhance the offloading rate. Then, given a specific volume of computational tasks to be offloaded, the offloading delay can be reduced. Furthermore, RISs are also capable of contributing to the reduction of the computing delay. Specifically, the volume of data offloaded to MEC nodes is critically dependent on the transmission rate of offloading links, in a given transmission time slot. As a benefit of the aforementioned reflection-based virtual link, the assistance of RISs allows more data to be offloaded. Hence, the data can be processed at the MEC node in a more prompt manner than by local computing at the vehicles, also taking into account the data/information provided by other vehicles. 

\subsubsection{Improving Energy Efficiency}
In MEC systems, the devices' energy efficiency is defined as the ratio of the total number of bits processed to the sum of all devices' energy consumption of both the local computing and of the computational offloading, which is of paramount importance for the IoT devices equipped with small batteries. 
As indicated in Fig.~\ref{fig:use_case}, the deployment of RISs is capable of substantially improving the devices' energy efficiency. Specifically, RISs are capable of attaining reflection-based passive beamforming gain for the computational offloading links, by proactively adjusting the phase shift of the signal reflected. They can also achieve beneficial diversity gain by combining all the signals reflected by the individual RIS elements. Then, compared to the system operating without RISs, a certain target offloading rate can be attained by using a reduced offloading power in RIS-assisted MEC systems.
Hence, given a specific volume of computational tasks to be offloaded, the transmit duration can be reduced, thus reducing the energy consumption.
Furthermore, the emerging technique of energy harvesting enables wireless devices to be charged wirelessly. Hence, in \cite{bai2020resource} an innovative wirelessly powered MEC architecture has been proposed for prolonging the network life, where the devices' batteries are replenished relying on wireless energy transfer from hybrid APs in the first stage, followed by computational offloading in the second stage. Given that the aforementioned RIS-gains can be exploited in both stages, the energy efficiency of the hybrid AP can also be improved. 
Bearing in mind that the deployment of a RIS increases the energy consumption owing to the excessive coordination among the AP, the MEC node and a RIS, the energy consumption is inevitably increased. However, compared to that of data transmission, the energy consumption of overheads may still remain less dominant.

\subsubsection{Enhancing the Task-Completion Rate} 
The total task completion rate is defined as the total number of bits that can be processed in a time slot, given a limited energy budget. This plays a crucial role in indicating whether the system considered is capable of supporting computation-intensive applications, such as face/speech recognition and big data analysis. 
In the conventional MEC system, although MEC nodes are avaiable in the vicinity of devices for providing computation services, a large fraction of computational resources available in the MEC nodes often remain idle, because local computing is used instead of computational offloading, when the offloading links are imperfect. As shown in Fig.~\ref{fig:use_case}, upon imposing the wireless propagation environment relying on RISs, more data can be offloaded to MEC nodes, whose powerful computing capabilities can be exploited for supporting the aforementioned computationally intensive collaborative applications \cite{chu2020intelligent}.

\subsubsection{Secure Computation Offloading}

In IoT networks, the data offloaded to MEC nodes usually contains environmental information or clues concerning personal actions, which may also be of valuable to third parties. Given that wireless signal propagation can be accessed both by authorized users and by adversaries, the offloading links are subject to eavesdropping, which jeopardizes data security and privacy \cite{bai2019energy}. Furthermore, the important information, such as equipment ID, communication protocol, and the address of wireless nodes can also be observed by adversaries. To overcome these security issues, the reflection coefficients of the RIS can be jointly designed with the multi-user detector (MUD) weights of the AP for enhancing the legitimate wireless devices' signal-to-interference-plus-noise ratio (SINR). As part of this joint design, the SINR at the eavesdropper may also be simultaneously reduced by jointly designing the reflection coefficients of the RIS together with a jammer beam for transmitting artificial noise to drown out the eavesdropper. Both schemes or their combinations are capable of enhancing the secrecy capacity of the offloading link \cite{lu2019intelligent}.

\subsection{Challenges and Solutions}

Deploying RISs in MEC systems imposes new challenges on its system design, which are discussed below along with their potential solutions.

\subsubsection{RIS Reflection Coefficient Design}

The RIS reflection coefficient optimization problem is typically a non-convex quadratically constrained quadratic program subject to the unit-modulus constraint imposed on the reflection phase shifts, which can be readily solved by using the conventional semi-definite relaxation technique. However, this incurs high complexity owing to the high number of randomizations required for approaching the optimal value. As a compromise, a locally optimal solution can be found with the aid of low-complexity iterative algorithms, such as the majorization-minimization algorithm, the complex circle manifold method, and successive convex optimization. Furthermore, given that the channel state information (CSI) of each reflective element in RIS-assisted systems is not always available owing to its potentially excessive channel-estimation complexity, both powerful robust optimization techniques and machine learning aided optimization \cite{huang2020reconfigurable} can be applied. 

\subsubsection{Joint Communications and Computation Optimization}

Since diverse applications are supported by MEC systems, it is imperative to carefully adapt both the communications and computational resource allocation to their heterogeneous requirements. Specifically, the communication resource allocation includes their multiple access schemes, spectral and temporal resource allocation to wireless devices, RIS configuration, and the design of the MUD matrix at the AP. By contrast, the computational resource allocation entails the offloading decisions, the design of the central processing unit (CPU) clock frequencies of wireless devices, and the computational resource allocation at the MEC nodes \cite{zhou2020delay}.
Since these diverse variables are optimized in this joint system design, the optimization problem cannot be solved using a single optimization technique. Hence, a combination of convex and of non-convex optimization techniques have to be invoked. 
However, if the optimization problem formulated is NP-hard, game theory, genetic algorithms, bio-inspired or machine learning techniques can be invoked.

\section{Case Study: Exploiting RISs for Shortening Processing Latency in MEC Systems}

\begin{figure}[t!]
\centering
\includegraphics[width=3.5in]{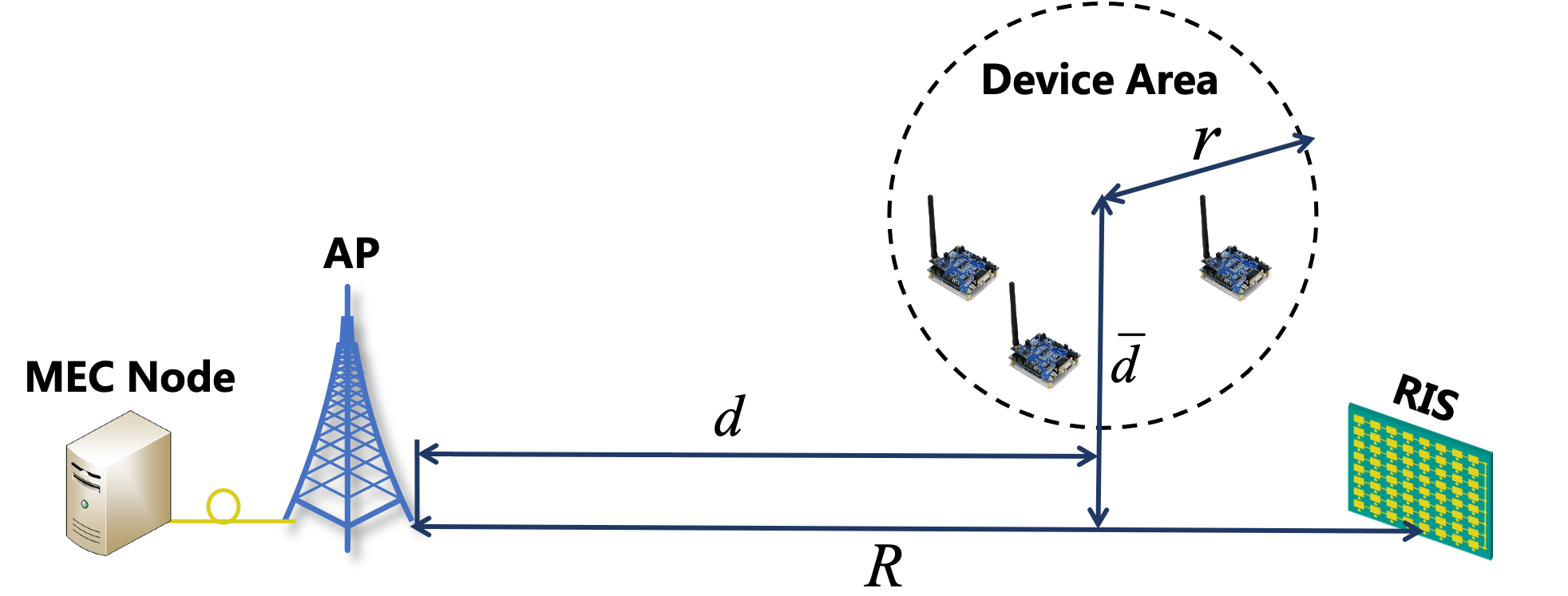}
\caption{Top view of the locations of the RIS-assisted MEC systems for simulation settings.}
\label{fig:simulation}
\end{figure}

\begin{figure*}[t!]\centering
\subfloat[]{\includegraphics[width=3.6in]{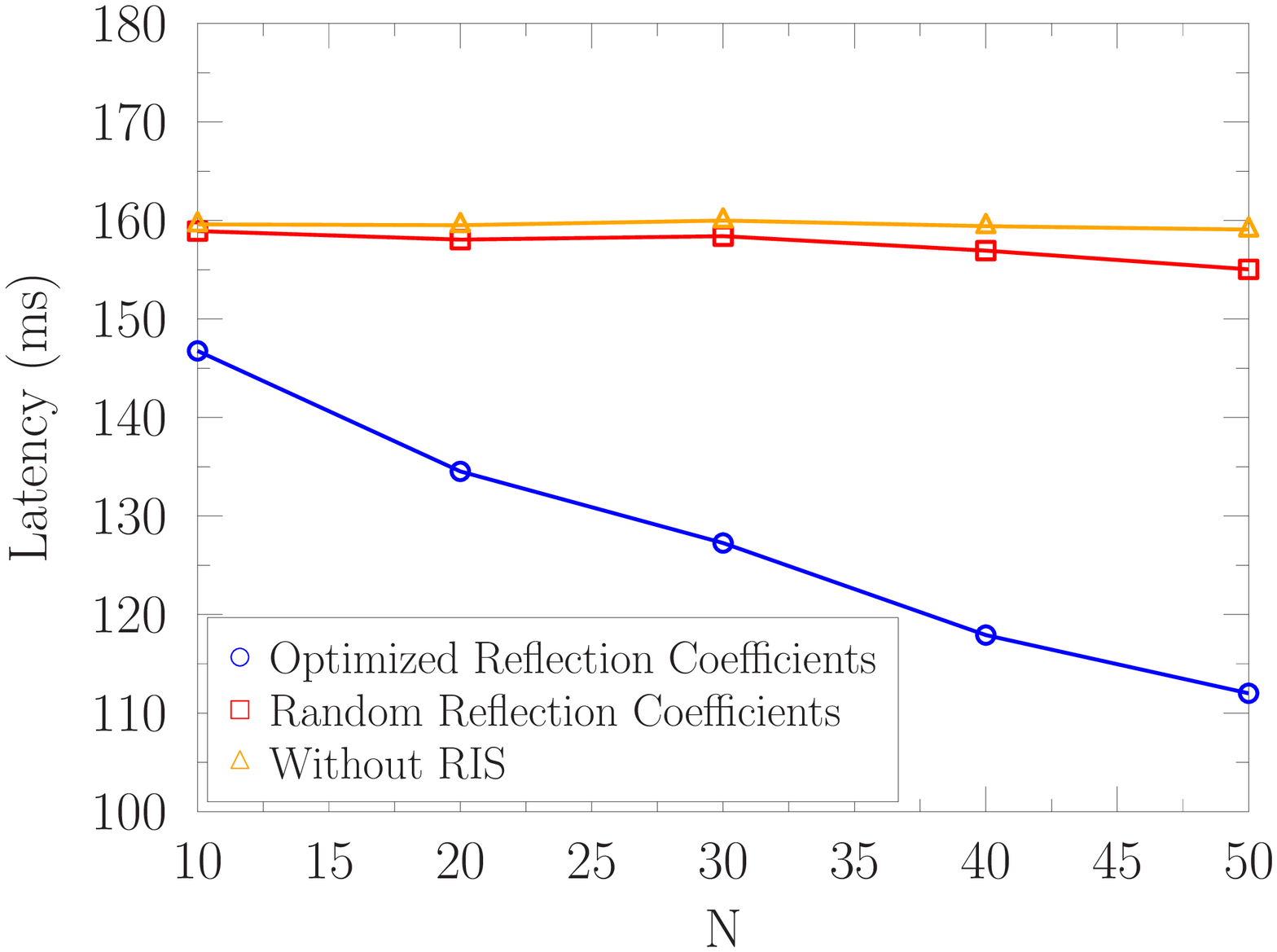}\label{fig:MU_N}}
\subfloat[]{\includegraphics[width=3.6in]{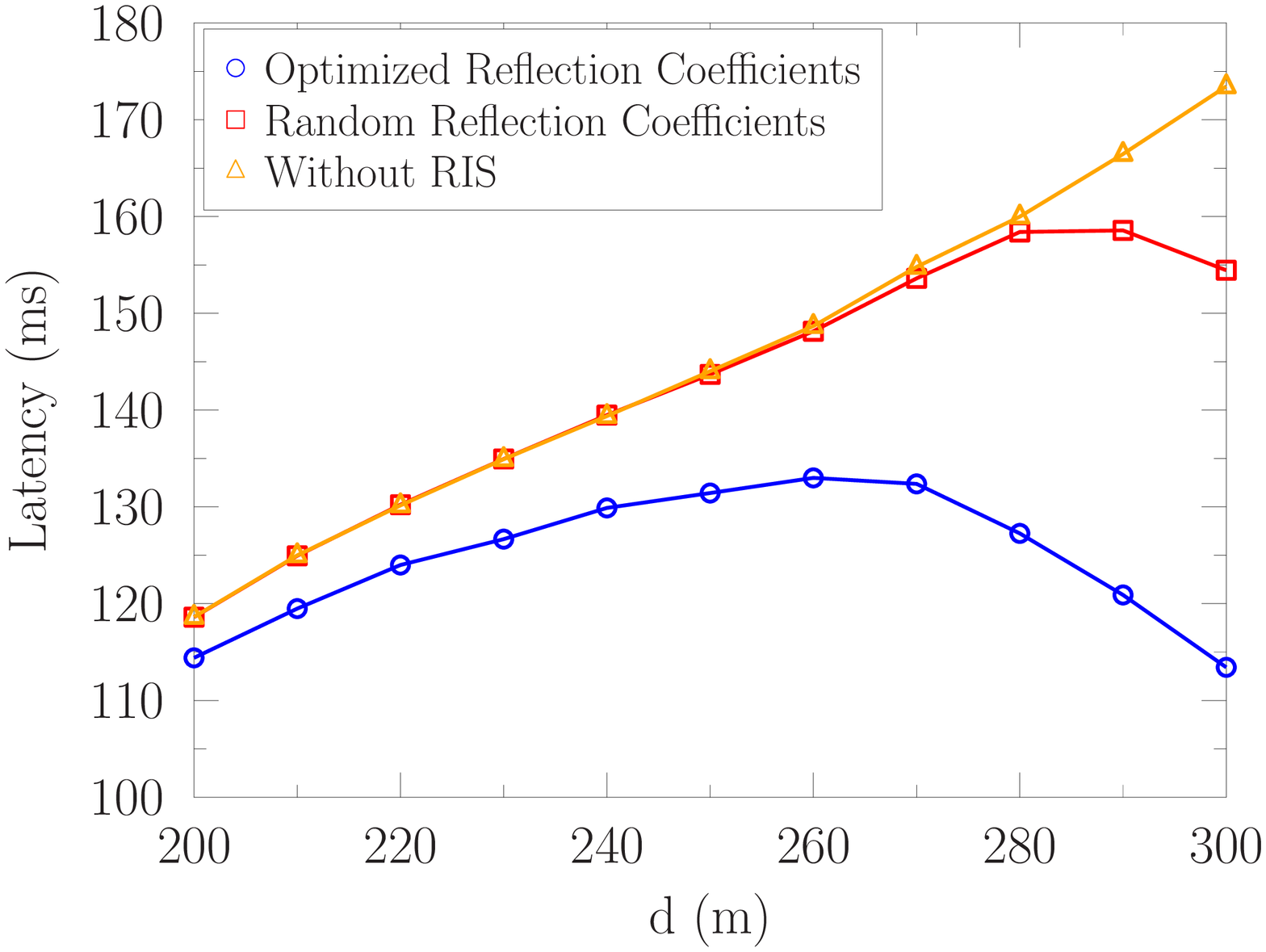}\label{fig:MU_d}}
\caption{Simulation results of the latency versus the number of reflection elements and the distance between devices and the AP. (a): $d = 280~\rm{m}$; (b): $N = 30$.}
\label{fig:MU}
\end{figure*}

As shown in Fig.~\ref{fig:simulation}, we consider a RIS-assisted MEC system where $K$ single-antenna wireless devices may opt for offloading their computational tasks to the MEC nodes via an $M$-antenna AP with the aid of a RIS having $N$ reflection elements. 
The distance between the RIS and the AP is $R$, while the wireless devices are within a circle of a radius $r$.
It is assumed that the direct device-AP link is obstructed while the location of the RIS has been carefully selected for attaining near-LoS propagation for both the device-RIS and the RIS-AP channels. This may be readily arranged by partitioning both the RIS and the AP above the level of the local paraphernalia.
Hence, we assume that the direct device-AP channel follows Rayleigh fading having a path-loss exponent of $3.5$, while the device-RIS and RIS-AP channels are dominated by the LoS link associated with a path-loss exponent of $2.2$. The computation offloading takes place over a given frequency band of $1~\rm{MHz}$ in the same time resource. The total number of bits to be processed varies in the range of $250~\rm{Kb}$ to $350~\rm{Kb}$. The maximum clock rate of the MEC node is $50\rm{GHz}$.
Both the communications resource including the MUD matrix and the RIS reflection-coefficients as well as the computing resources such as the volume to be offloaded and CPU clock frequencies at the MEC nodes are jointly optimized for reducing the processing delay.
In order to demonstrate the benefit of employing a RIS in MEC systems, we compare the following three schemes under the setup of $M=5$,  $K=4$, $R=300~\rm{m}$ and $r=10~\rm{m}$. 
\begin{itemize}
\item \emph{Without RIS:} The computational tasks are offloaded via the direct LoS device-AP link. The computation and communications designs are optimized alternatively. Specifically, both the volume of the computation offloading and edge computing resource allocation are optimized based on convex optimization, while the MUD matrix is optimized with the aid of the minimum mean square error (MMSE) method.
\item \emph{Random Reflection Coefficients:} The computational offloading takes place with the aid of both the direct device-AP and of the reflection-based device-RIS-AP link. Specifically, both the volume of the computation offloading and edge computing resource allocation are optimized based on convex optimization, while the MUD matrix is optimized with the aid of the minimum mean square error (MMSE) method and the RIS reflection coefficients are set as random values.
\item \emph{Optimized Reflection Coefficients:} The computation offloading takes place with the aid of both the direct device-AP and of the reflection-based device-RIS-AP link. Specifically, both the volume of the computation offloading and edge computing resource allocation are optimized based on convex optimization, while the MUD matrix is optimized with the aid of the MMSE method and the RIS reflection coefficients are optimized relying on the majorization-minimization algorithm.
\end{itemize}

As mentioned in Section~\RN{2}-C1, the processing delay in MEC systems is the sum of the delay induced by computation offloading, by computing at MEC nodes, and by computation feedback. Since the computation feedback is usually of a small volume, we neglect feedback duration.
Fig.~\ref{fig:MU_N} shows the simulation results of the processing latency versus the number of reflective elements. Observe that the performance gap between the schemes of ``Without RIS" and of ``RandPhase" increases upon increasing the number of reflective elements, which implies that RISs are capable of enhancing the performance of MEC systems even without optimizing the RIS reflection-coefficients responses. This is mainly due to the diversity gain of combining all the signal by elements. Furthermore, the performance gap between the schemes of ``With RIS" and of ``RandPhase" is $12~\rm{ms}$ when $N=10$, while it becomes $43~\rm{ms}$ when $N=50$, which implies that a higher number of reflective elements leads to an increased beamforming gain. Explicitly, RISs are capable of substantially reducing the processing latency of MEC systems.
Fig.~\ref{fig:MU_d} plots the simulation results of the processing latency versus the distance between wireless devices and the AP. Observe that the advantage of deploying RISs gradually increases, when the distance between the wireless devices and the RIS becomes smaller, where the device-RIS-AP link dominates the computation offloading, which confirms that a higher gain can be achieved in the near-RIS area.

\section{Future Research Opportunities}

\subsection{RIS Deployment in MEC Systems}

When designing a MEC system, the distribution of MEC servers, of APs, and of wireless devices should be taken into consideration for beneficially deploying the RISs. However, a fraction of APs have their own MEC servers and the different MEC servers typically have diverse computational capabilities. The RIS locations have to be carefully decided for striking a trade-off between the queuing delay caused by the wireless traffic specific at the APs having their co-located MEC servers and that by the task scheduling in the backhaul connecting the MEC servers and those APs which do not have any computational resources. 
On the other hand, the computation demands are normally non-uniformly distributed, hence they don't lend themselves to modeling by the usage of the conventional homogenenous Poisson point processes (HPPP) when planning the locations of both MEC servers and of RISs. As a possible solution, Ginibre $\alpha$-determinantal point process may be used for characterizing their positions \cite{mao2017survey}.

\subsection{Joint Communications and Computation Optimization for Heterogeneous Applications and Wireless Devices}

Given a diverse variety of applications and of the wireless devices used in RIS-assisted MEC systems, separate communications and computation designs focusing on a specific requirement are incapable of meeting all their demands. Therefore, a joint communications and computation design should be conceived. In this context, the associated applications should control the wireless devices' offloading decisions, which directly affects the communications resource management policy. Secondly, the different applications impose diverse QoE requirements on MEC systems. For example, smart manufacturing exhibits stringent delay specifications, but loose energy consumption requirements, whereas the MEC services designed for health monitoring are expected to accommodate both energy consumption and offloading security requirements. In this case, how to jointly optimize both RIS reflection coefficients as well as the communications and computation resource allocation remains an open issue, calling for the systematic exploration of the entire Pareto front of optimal solutions.

\subsection{On-Demand Device-RIS-Edge Association}

Device association in conventional cellular networks, which generally opts for the specific AP associated with the maximum signal-to-interference-plus-noise ratio (SINR), cannot be directly applied in RIS-assisted MEC systems, for the following reasons. Firstly, owing to their intrinsic beamforming capabilities, RISs are capable of boosting the SINR of the link between a wireless device and a particular AP, especially in the emerging cell-free systems. This capability introduces a new degree-of-freedom in network association. Secondly, the SINR values of offloading links only decide upon the computation offloading rate, while the wireless devices' computation offloading decisions are also jointly determined by the applications operated, by wireless devices' computational capabilities and battery lives, as well as by the idle/busy states of MEC servers. Thirdly, in practice, the volume of computational tasks may vary extensively over time, leading to a dynamically fluctuating processing delay or energy consumption at the wireless devices. Therefore, it is imperative to address the computational task dynamics by conceiving an on-demand device-RIS-edge association scheme. A possible solution is to rely on the user-centric network association strategy of \cite{chen2016user} while additionally considering the specific role of RISs.

\subsection{Specific System Design for Coexistence of Offloading Data and Conventional Cellular Data}

Owing to their ubiquitous coverage, cellular networks may be relied upon for computation offloading in MEC systems. However, this may drastically reconfigure the conventional uplink/downlink tele-traffic distribution. Specifically, downlink wireless traffic is the dominant direction in the conventional cellular networks, while MEC systems mainly rely on uplink transmission. Therefore, in order to support the coexistence of offloading data and of conventional cellular data, it is imperative to conceive sophisticated full-duplex systems.
Fortunately, as a benefit of their intrinsically passive reflective nature, RISs are ideally suited for enhancing both the uplink and downlink transmission in a simultaneous manner by collaborating with full-duplex APs. However, we have to bear in mind that in contrast to cellular users whose QoE is only dependent on the communications quality, the QoE of MEC users is dependent both on the computation resource allocation at the MEC nodes and on the wireless devices' specific features, e.g. their local computational capabilities and battery charges. Hence, the joint communications and computation design necessitates a profound redesign of the system for fully exploiting the beneficial features of RISs when jointly supporting the offloading data and conventional cellular data.

\section{Conclusions}

In this article, we discussed the potential of RISs in improving the MEC systems in terms of processing latency, energy efficiency, total task completion rate, and secure computation offloading. The benefits of RISs were quantified by exemplifying their processing delay reduction attained in MEC systems. Since RIS-assisted MEC systems have hitherto remained largely unexplored, we also provided a range of future research ideas.

\appendices

\ifCLASSOPTIONcaptionsoff
  \newpage
\fi

\bibliographystyle{ieeetr}
\bibliography{IEEEabrv.bib}

\begin{IEEEbiographynophoto}
{Tong Bai}
(S'15–M'19) received the Ph.D. degrees in communications and signal processing from the University of Southampton in 2019. Then, he held the position of Postdoctoral Researcher in Queen Mary University of London. Since 2020, he has been with Beihang University (BUAA) as an Assistant Professor. His research interests include the performance analysis, transceiver design, and utility optimization for wireline and wireless communications.
\end{IEEEbiographynophoto}

\begin{IEEEbiographynophoto}
{Cunhua Pan}
received the Ph.D. degrees from Southeast University, Nanjing, China, in 2015. From 2015 to 2019, he was a Research Associate at the University of Kent and later in Queen Mary University of London, U.K., where he is currently a Lecturer.
His research interests mainly include reconfigurable intelligent surface, machine learning, UAV, Internet of Things, and mobile edge computing.
He serves as an Editor of IEEE WIRELESS COMMUNICATION LETTERS.
\end{IEEEbiographynophoto}

\begin{IEEEbiographynophoto}
{Chao Han} (S’18) is currently working toward the Ph.D. degree in communication and information systems with the School of Electronic and Information Engineering, Beihang University, Beijing, China. His research interests include signal processing both for wireless communications and for Internet of Things.
\end{IEEEbiographynophoto}


\begin{IEEEbiographynophoto}
{Lajos Hanzo}  (F’04) FREng, FIEEE, FIET, Fellow of EURASIP, DSc, received his degree in electronics in 1976 and his doctorate in 1983. He holds an honorary doctorate from the Technical University of Budapest (2009) and from the University of Edinburgh (2015). He is a member of the Hungarian Academy of Sciences and a former Editor-in-Chief of the IEEE Press. He is a Governor of both IEEE ComSoc and of VTS.
\end{IEEEbiographynophoto}

\end{document}